\documentclass[12pt]{JHEP3}

\usepackage{epsfig}

\def\spose#1{\hbox to 0pt{#1\hss}}
\def\ltapprox{\mathrel{\spose{\lower 3pt\hbox{$\mathchar"218$}}
 \raise 2.0pt\hbox{$\mathchar"13C$}}}
\def\gtapprox{\mathrel{\spose{\lower 3pt\hbox{$\mathchar"218$}}
 \raise 2.0pt\hbox{$\mathchar"13E$}}}

\title{
SU($N$) gauge theories in the presence of a topological term
}

\author{Luigi Del Debbio \\
        SUPA, School of Physics, University of Edinburgh, 
        Edinburgh, EH9 3JZ, UK\\
        E-mail: \email{luigi.del.debbio@ed.ac.uk} 
} 

\author{Haralambos Panagopoulos \\ 
        Department of Physics, University of Cyprus,
        Lefkosia, CY-1678, Cyprus\\ 
        E-mail: \email{haris@ucy.ac.cy} 
}

\author{Ettore Vicari \\ 
        Dipartimento di Fisica, Universit\`a 
        di Pisa, and INFN, I-56127 Pisa, Italy \\ 
        E-mail: \email{vicari@df.unipi.it} 
}

\abstract{ We review recent results on the $\theta$ dependence of the
  ground-state energy and spectrum of four-dimensional SU($N$) gauge theories,
  where $\theta$ is the coefficient of the CP-violating topological term
  $F\widetilde{F}$ in the Lagrangian.  In particular, we discuss the results
  obtained by Monte Carlo simulations of the lattice formulation of QCD, which
  allow the investigation of $\theta$ dependence around $\theta=0$ by
  determining the moments of the topological charge distribution, and their
  correlations with other observables.  The results for $N=3$ and larger
  values of $N$ support the scenario obtained by general large-$N$ scaling
  arguments.  }

\begin{document}

\section{$\theta$ dependence of the ground-state energy}
\label{gse}

Four-dimensional SU($N$) gauge theories have a nontrivial dependence on the
angle $\theta$ that appears in the Euclidean Lagrangian as
\begin{equation}
{\cal L}_\theta  = {1\over 4} F_{\mu\nu}^a(x)F_{\mu\nu}^a(x)
- i \theta {g^2\over 64\pi^2} \epsilon_{\mu\nu\rho\sigma}
F_{\mu\nu}^a(x) F_{\rho\sigma}^a(x)
\label{lagrangian}
\end{equation}
where $q(x)=\frac{g^2}{64\pi^2} \epsilon_{\mu\nu\rho\sigma} F_{\mu\nu}^a(x)
F_{\rho\sigma}^a(x)$ is the topological charge density. 
Semiclassical instanton solutions are one example of field
configurations which have nontrivial topological
properties. Moreover, the most plausible explanation of how the solution of
the so-called U(1)$_A$ problem can be compatible with the $1/N$ expansion
(performed keeping $g^2N$ fixed \cite{Hooft-74}) requires a nontrivial
$\theta$ dependence of the ground-state energy density
$F(\theta)$~\cite{Witten-79,Veneziano-79},
\begin{equation}
\exp[ - V F(\theta) ] = 
\int [dA] \exp \left(  - \int d^4 x {\cal L}_\theta \right)
\label{vftheta}
\end{equation}
where $V$ is the volume. Evidence for such a dependence has been 
obtained by exploiting the lattice formulation of the theory, using
numerical Monte Carlo simulations, as will be described in Section
\ref{fte}.  The complex nature of the $\theta$ term in the Euclidean
QCD Lagrangian makes the Monte Carlo studies of the $\theta$
dependence quite hard, since the lattice action corresponding to the
Lagrangian (\ref{lagrangian}) cannot be directly simulated for
$\theta\ne 0$.  Nevertheless, important information on the $\theta$
dependence of relevant physical quantities, such as the ground-state
energy and the spectrum, can also be inferred from results at
$\theta=0$, by expanding them about $\theta=0$ and computing the
coefficients of the expansion~\cite{DPV-02,DMPSV-06}.
The $\theta$ dependence is particularly interesting in the large-$N$
limit where the issue may also be addressed by other approaches, such
as AdS/CFT correspondence applied to nonsupersymmetric and non
conformal theories, see e.g. Ref.~\cite{AGMOO-00}.

We introduce a {\it scaling} energy density
\begin{equation}
f(\theta) = {\Delta F(\theta)\over \sigma^2},
\label{scge}
\end{equation}
where $\Delta F(\theta) \equiv F(\theta) - F(0)$ and $\sigma$ is the string
tension at $\theta=0$.  
By expanding $f(\theta)$ around $\theta=0$, one
can study its $\theta$ dependence in the region of small $\theta$
values.
The function $f(\theta)$ is conveniently
parametrized as
\begin{eqnarray}
f(\theta)={1\over 2} C \theta^2 s(\theta),\label{ftheta}
\end{eqnarray}
where $C$ is the ratio
$\chi/\sigma^2$ and $\chi$ is the topological susceptibility at $\theta=0$,
\begin{equation}
\chi = \int d^4 x \langle q(x)q(0) \rangle = {\langle Q^2 \rangle\over V}
\label{chidef}
\end{equation}
where $Q=\int d^4 x q(x)$.  $s(\theta)$ is a dimensionless function of
$\theta$ such that $s(0)=1$.  

The function $s(\theta)$ can be expanded around $\theta=0$ as 
\begin{eqnarray}
s(\theta) = 1 + b_2 \theta^2 + b_4 \theta^4 + \cdots.
\label{stheta}
\end{eqnarray}
The coefficients of the expansion of $f(\theta)$ are related to the
zero-momentum $n$-point connected correlation functions of the topological
charge density, and therefore to the moments of the probability distribution
$P(Q)$ of the topological charge $Q$.  If $s(\theta)=1$, 
and therefore $b_{2n}=0$,
the corresponding distribution $P(Q)$ is Gaussian, i.e.
\begin{equation}
P(Q)={1\over \sqrt{2\pi\langle Q^2\rangle}}\,{\rm exp}\left( -{Q^2\over 2\langle Q^2\rangle}
\right).
\end{equation}
Therefore the coefficients
$b_{2n}$ of the expansion of $s(\theta)$ parametrize the deviations from a
simple Gaussian behavior.  For example, he first non--trivial
correction is given by
\begin{eqnarray}
&&b_2 = - {\chi_4 \over 12 \chi},\label{b2chi4} \\
&& \chi_4 = {1\over V} \left[ 
\langle Q^4 \rangle_{\theta=0} - 3 \left( 
\langle Q^2 \rangle_{\theta=0} \right)^2 \right]. \label{chi4}
\end{eqnarray}
It has been recently shown~\cite{Luscher-04} (see also \cite{GRT-04})
that correlation functions  
involving multiple zero-momentum insertions of the topological charge density
can be defined in a nonambiguous, regularization-independent way, and
therefore the expansion coefficients $b_{2n}$ are well defined
renormalization-group invariant quantities.

\section{Behavior in the large-$N$ limit}
\label{largeN}

Witten argued \cite{Witten-80} that in the large-$N$ limit $F(\theta)$ is a
multibranched function of the type
\begin{equation}
F(\theta) = N^2 {\rm min}_k\, H\left( {\theta+2\pi k\over N}\right)
\label{conj1}
\end{equation}
which is periodic in $\theta$, but not smooth
since at some value of $\theta$ there is a jump between two different
branches. 
This issue was also discussed in Ref. \cite{Ohta}. More recently, the
conjecture was refined \cite{Witten-98} 
leading to a rather simple expression for $\Delta F(\theta)$ in the
large-$N$ limit, that is 
\begin{equation}
\Delta F(\theta) = {\cal A} \, {\rm min}_k \, (\theta+2\pi k)^2 + O\left(
1/N\right).
\label{conj2}
\end{equation} 
In particular, for sufficiently small values of $\theta$,
i.e. $|\theta|<\pi$, 
\begin{equation}
\Delta F(\theta) = {\cal A} \, \theta^2  + O\left( 1/N\right).
\label{conj2b}
\end{equation} 
Thus possible $O(\theta^4)$ terms are expected to be depressed by
powers of $1/N$.  This conjecture has been supported using arguments
based on duality between large-$N$ gauge theories and string theory
\cite{Witten-98}.  It has also been discussed in a field-theoretical
framework in Ref.~\cite{Gabadadze-99}.

The large-$N$ behavior of the coefficients $b_{2n}$ of
the expansion of $f(\theta)$ around $\theta=0$
can be inferred by using general large-$N$ scaling arguments applied to the
Lagrangian (\ref{lagrangian}). They indicate the ratio $\bar{\theta}\equiv
\theta/N$ as the relevant quantity in the large-$N$ limit of the ground-state
energy, and more generally of the spectrum of the theory.  Then we expect
\begin{eqnarray} 
&&f(\theta) = N^2 \bar{f}(\bar{\theta}\equiv \theta/N),
\label{fthetabar} \\
 &&\bar{f}(\bar{\theta}) = 
{1\over 2} C_\infty \bar{\theta}^2 ( 1 + \bar{b}_2 \bar{\theta}^2 + 
 \bar{b}_4 \bar{\theta}^4 + \cdots), 
\label{lnexp}
\end{eqnarray}
where $C_\infty$ is the large-$N$ limit of the ratio $C=\chi/\sigma^2$.
Comparing with Eq.~(\ref{ftheta}), one derives
\begin{eqnarray}
C=C_\infty + c_2/N^2 + ... , \qquad b_{2i}=\bar{b}_{2i}/N^{2i}+...,
\label{largeNco}
\end{eqnarray}
We recall that a nonzero value of
$C_{\infty}$ is essential to provide an explanation to the U(1)$_A$
problem in the 't~Hooft large-N limit, and can be related to the $\eta'$ mass
\cite{Witten-79,Veneziano-79} through the relation
\begin{equation}
\chi_\infty = {f_\pi^2 m_{\eta'}^2\over 4 N_f} + O(1/N).
\label{wittenformula}
\end{equation}
The quantity $b_2$ also lends itself to a physical interpretation,
being related to the $\eta^\prime - \eta^\prime$ elastic scattering
amplitude \cite{Veneziano-79}.

\section{Results for the first few terms of the expansion 
around $\theta=0$ of the ground-state energy}
\label{fte}

The $\theta$ dependence of SU($N$) gauge theories has been investigated by
Monte Carlo simulations of their Wilson lattice formulation. The
lattice action corresponding 
to the Lagrangian (\ref{lagrangian}) cannot be directly simulated for
$\theta\ne 0$, by virtue of the complex nature of the $\theta$
term. On the other hand, 
the coefficients $b_{2n}$ in the expansion
of the ground-state energy $F(\theta)$ around $\theta=0$ can
be accessed by determining the moments of the topological charge
distribution at $\theta=0$.
They are dimensionless renormalization-group invariant
quantities, which should approach a constant in the continuum limit, with
$O(a^2)$ scaling corrections ($a$ is the lattice spacing).

Computing quantities related to topology using lattice simulation techniques
is not a simple task.  In the case $N=3$ several methods have been employed to
determine the topological susceptibility, see e.g. Refs.
\cite{DFRV-81}-\cite{HIMT-01},\cite{DPV-02},\cite{LTW-05}-\cite{DFHK-07}.
Cooling, geometrical, heating techniques have been used to address the
problems caused by power--divergent additive contributions and multiplicative
renormalizations in definitions of the topological susceptibility based on
discretized versions of the topological charge density operator $q(x)$.  These
methods have their drawbacks, since their systematic errors are not under
robust theoretical control.

A substantial progress has been achieved after the introduction of the
Neuberger overlap formulation~\cite{Neuberger-98,Neuberger-01} of fermions,
which represented a breakthrough for the lattice formulation of QCD.  Overlap
lattice fermions satisfy the Ginsparg-Wilson relation~\cite{GW-82} and
therefore preserve an exact chiral symmetry~\cite{Luscher-98}.  As a by
product, the index of the overlap Dirac operator \cite{HLN-98}
provides a well--defined estimator for the topological 
charge~\cite{Neuberger-01,GRTV-01,GRT-04,Luscher-04},  which can also
be used in pure 
gauge theories. This method circumvents completely the problem of
renormalization arising in bosonic approaches, even though at a much
higher computational cost. Using these methods, the topological
susceptibility of the pure SU(3) gauge theory has been investigated in
Refs.
\cite{EHN-98}-\cite{DGP-05},
finally obtaining the accurate estimate~\cite{DGP-05} $\chi r_0^4=0.059(3)$
($r_0$ is the length scale defined in~\cite{Sommer}).  This value corresponds
to $C=\chi/\sigma^2=0.029(2)$ (using~\cite{NRW-01} $\sigma^{1/2}
r_0=1.193(10)$).

It is important to note that the results obtained by the
(less computer-power demanding) bosonic methods are substantially consistent,
see e.g.  Refs.~\cite{ADD-97,Teper-00,LT-01,DPV-02,DP-04,DFHK-07}, showing
their effectiveness although they are supported by a weaker theoretical
ground.  For example, we mention the results:~\cite{ADD-97} $C=0.027(4)$,
obtained using the heating method, ~\cite{DPV-02} $C=0.0282(12)$, obtained
using cooling, and the more recent result \cite{DFHK-07} $C=0.0259(11)$\,.

For larger values of $N$, results have been obtained only by the
cooling method so far~\cite{LT-01,DPV-02,LTW-05}, up to $N=8$.  They fit
well the expected large-$N$ behavior: $C=C_\infty+c_2/N^2$, providing
an estimate of $C_\infty$, and therefore of the topological
susceptibility in the large-$N$ limit: 
$C_\infty=0.0200(43)$ ~\cite{LT-01}, 
$C_\infty=0.0221(14)$ ~\cite{DPV-02}, 
$C_\infty=0.0248(18)$ ~\cite{LTW-05} (the latter was obtained using
$N\le 8$ and keeping $a$ fixed). These results are in substantial
agreement with 
the large-$N$ relation (\ref{wittenformula}).  We stress that the good
agrement for $N=3$ of the cooling method with the more rigorous
overlap result make us quite confident on the reliability of results
for higher values of $N$, since there are no arguments to suggest that
this agreement could be spoiled with increasing $N$ (actually there
are reasons in favor of improved agreement \cite{CTW-02,RRV-97}).  An
independent determination of $C_\infty$ using other methods would be
welcome.

Higher moments of the topological charge distribution provide estimates of the
coefficients $b_{2n}$ of the expansion of the scaling energy density
$f(\theta)$, cf. Eqs.~(\ref{ftheta}) and (\ref{stheta}).  
In particular $b_2$ can be estimated
using formulae (\ref{b2chi4}, \ref{chi4}).  There are a number of results at
$N=3$, obtained by different approaches: Ref.~\cite{DPV-02} used the cooling
method, Ref.~\cite{Delia-03} used the heating technique to estimate additive
and multiplicative renormalizations in zero-momentum correlations of lattice
discretizations of $q(x)$, and finally Ref.~\cite{GPT-07} used the most
rigorous and CPU intensive overlap method.  
The results reported in Table~\ref{b2} are in good agreement, suggesting that the
systematic errors of the various methods are sufficiently small. 
We mention that the fourth moment
of the topological charge distribution has been numerically investigated also
in Ref.~\cite{DFHK-07}, without arriving at any definite conclusion.  

The results of Table~\ref{b2} provide robust evidence that $b_2$ is nonzero,
and therefore that there are deviations from a Gaussian distribution of the
topological charge.\footnote{ An apparently contradictory result has been
  reported in Refs.~\cite{BCNW-03,GLWW-03,DGP-05} for the expected
  large-volume probability distribution $P(Q)$, i.e.  $P(Q)=(2\pi\langle
  Q^2\rangle)^{-1/2}e^{-{Q^2\over 2\langle Q^2\rangle}}\left[1
    +O(1/V)\right]$.  
A purely gaussian behaviour would imply an exact 
quadratic form for $f(\theta)$, and in particular a vanishing $b_2$,
thereby contradicting  
the assumption of a generic expansion of $f(\theta)$.}
However, $b_2$
turns out to be quite small, indeed $|b_2|\ll 1$. Thus deviations from a
simple Gaussian behavior are already small at $N=3$.

\TABLE[ht]{
\caption{ 
Results for the coefficient $b_2$ of the expansion (\ref{stheta})
}
\label{b2}
\begin{tabular}{cccl}
\hline\hline
\multicolumn{1}{c}{$N$}&
\multicolumn{1}{c}{Ref.}&
\multicolumn{1}{c}{method}&
\multicolumn{1}{c}{$b_2$}\\
\hline \hline
3  &  \cite{DPV-02} & cooling & $-$0.023(7) \\  
   &  \cite{Delia-03} & heating & $-$0.024(6) \\  
   &  \cite{GPT-07} & overlap & $-$0.025(9) \\  
4  &  \cite{DPV-02} & cooling & $-$0.013(7) \\  
6  &  \cite{DPV-02} & cooling & $-$0.01(2) \\  
\hline\hline
\end{tabular}
}

There are also estimates for
larger values of $N$, see Table~\ref{b2}, but only using the cooling
method. Again, given the agreement found at $N=3$, higher $N$ results
should be sufficiently reliable.  They appear to decrease consistently
with the expectation from the large-$N$ scaling arguments, i.e.
$b_2\approx
\bar{b}_2/N^2$ with $\bar{b}_2\approx -0.2$.

We also mention that the analytical properties at $\theta=0$ have been recently
discussed and numerically checked in Ref.~\cite{ADD-05}.

\eject
Overall, these results support the scenario obtained by general large-$N$
scaling arguments, which indicate $\bar{\theta}\equiv \theta/N$ as the
relevant Lagrangian parameter in the large-$N$ expansion.  They also
show that $N=3$ is already in the regime of the large-$N$ behavior.
For $N\ge 3$ the simple Gaussian form 
\begin{equation}
F(\theta) \approx {1\over 2}\chi \theta^2
\label{gauform}
\end{equation}
provides a good approximation of the dependence on $\theta$ for a
relatively large range of values of $\theta$ around $\theta=0$.

\section{$\theta$ dependence at finite temperature}
\label{gseft}

Another interesting issue concerns the behavior of topological properties at
finite temperature, and in particular their change at the finite-temperature
deconfining transition, which is first order for $N\ge 3$, see e.g.
Ref.~\cite{LTW-04} and references therein.  This issue has been investigated
in a number of numerical works, see e.g.
Refs.~\cite{Teper-88,ADD-97,DGHS-98,GHS-02,DPV-04,LTW-05}, using different
methods. They show that the topological properties, and in particular the
topological susceptibility $\chi$, vary very little up to $T\lesssim T_c$.
They change across the transition, where $\chi$ shows a significant decrease.
Then, at high temperature $T\gg T_c$, where the instanton calculus is
reliable, a rather different scenario emerges~\cite{Gross:1980br}.  

Concerning the large-$N$ behavior (investigated by performing
simulations at various values of $N\ge 3$ \cite{DPV-04,LTW-05}), the
results indicate that $\chi$ has a 
nonvanishing large-$N$ limit for $T<T_c$, as at $T=0$, and that the
topological properties, and therefore $F(\theta)$, remain substantially
unchanged in the low-temperature phase, up to $T_c$.  On the other hand, above
the deconfinement phase transition, for $T>T_c$, $\chi$ shows a large
suppression, hinting at a vanishing large-$N$ limit for $T>T_c$.  These
results support the hypothesis put forward in Ref.~\cite{KPT-98}: At large $N$
the topological properties in the high-temperature phase, for $T>T_c$, are
essentially determined by instantons that are exponentially suppressed, i.e.
behave as $e^{-N}$, and therefore the topological susceptibility gets rapidly
suppressed in the large-$N$ limit.

\section{$\theta$ dependence of the spectrum}
\label{spectrum}

Another interesting issue concerns the $\theta$ dependence of the
spectrum of the theory. The analysis  of 
the $\theta$ dependence of the glueball spectrum using AdS/CFT
suggests that the only effect of the $\theta$ term in the leading
large-$N$ limit on the lowest spin-zero glueball state is that this
state becomes a 
mixed state of $0^{++}$ and $0^{-+}$ glueballs, as a consequence of
the fact that the $\theta$ term breaks parity, but its mass does not
change \cite{GI-04}.

Ref.~\cite{DMPSV-06} presented an exploratory numerical study of the
$\theta$ dependence in the spectrum of SU($N$) gauge theories. Again
numerical simulations of the Wilson lattice formulation were employed
to investigate the $\theta$ dependence of the string tension
$\sigma(\theta)$ and the lowest glueball mass $M(\theta)$. Around
$\theta=0$ one can write
\begin{eqnarray}
&&\sigma(\theta) = \sigma \left( 1 + s_2 \theta^2 + ... \right),
\label{sigmaex}\\
&&M(\theta) = M\left( 1 + g_2 \theta^2 + ... \right)
\label{gmex}
\end{eqnarray}
where $\sigma$ and $M$ are respectively the string tension and the
$0^{++}$ glueball mass at $\theta=0$.  Then the coefficients of these
expansions can be computed from appropriate correlators at $\theta=0$.
In particular, $s_2$ can be determined~\cite{DMPSV-06} from
the large-$t$ behavior of connected correlation functions of two
Polyakov lines at distance $t$ and the square topological charge, such
as
\begin{eqnarray}
\langle A_P(t) Q^2 \rangle_{\theta=0} -
\langle A_P(t) \rangle_{\theta=0} \langle Q^2 \rangle_{\theta=0}
\label{g2}
\end{eqnarray}
where
\begin{equation}
A_P(t) = \sum_{x_1,x_2}  {\rm Tr}\,P^\dagger(0;0) \; {\rm Tr}\,P(x_1,x_2;t),  
\label{apdef}
\end{equation}
$P(x_1,x_2;t)$ is the Polyakov line along the $x_3$ direction of size $L$, and
$Q$ is the topological charge.  Analogously, the $O(\theta^2)$ term of the
glueball mass can be obtained from appropriate connected correlation functions
of plaquette operators and $Q^2$. The $O(\theta^2)$ coefficients $s_2$ and
$g_2$ are dimensionless scaling quantities, which should approach a
constant in the 
continuum limit, with $O(a^2)$ scaling corrections.  

Ref.~\cite{DMPSV-06}
obtained the first estimates of $s_2$ and $g_2$ using the cooling method to
determine the topological charge, and for $N=3,4,6$ to also check their
large-$N$ behavior.
The $O(\theta^2)$ terms in the expansion around $\theta=0$ of the
spectrum of SU($N$) gauge theories are small for all $N\ge 3$,
especially when dimensionless ratios are considered, such as
$M/\sqrt{\sigma}$ and, for $N>3$, the ratios of independent $k$
strings.  For example we mention the estimates
$s_2=-0.08(1)$ and $g_2=-0.06(2)$ for $N=3$.  
One may also consider the $\theta$ dependence of the scaling ratio
\begin{equation}
{M(\theta)\over \sqrt{\sigma(\theta)}} =
{M\over \sqrt{\sigma}} ( 1 +  c_2 \theta^2 + ... ),
\label{ratioex}
\end{equation}
where $c_2=g_2-s_2/2$, thus $c_2=-0.02(2)$ for $N=3$. Moreover, the
$O(\theta^2)$ corrections appear to decrease with increasing $N$, and
the coefficients do not show evidence of convergence to a nonzero
value.  This is suggestive of a scenario in which the $\theta$
dependence of the spectrum disappears in the large-$N$ limit, at least
for sufficiently small values of $\theta$ around $\theta=0$.  In the
case of the spectrum, the general large-$N$ scaling arguments of
Sec.~\ref{largeN}, which indicate $\bar{\theta}\equiv \theta/N$ as the
relevant Lagrangian parameter in the large-$N$ limit, imply that
$O(\theta^2)$ coefficients should decrease as $1/N^{2}$.  The results
of Ref.~\cite{DMPSV-06} appear substantially consistent: In the case
of the string tension they suggest $s_2\approx -0.9/N^2$.  

Of course, further investigation is required to put this scenario on a firmer
ground, using for example other definitions of topological charge.

\section{The case of the two-dimensional CP$^{N-1}$ model}
\label{cpn}

Issues concerning the $\theta$ dependence can also be discussed in
two-dimensional CP$^{N-1}$ models~\cite{DDL-79,Witten-79b},
\begin{equation}
{\cal L} = {N\over 2g} \overline{D_\mu z}\, D_\mu z 
\label{lagrangiancpn}
\end{equation}
where $z$ is a $N$-component complex scalar field subject to the
constraint $\bar{z}z=1$, $A_\mu=i\bar{z}\partial_\mu z$ is a composite
gauge field, and $D_\mu =\partial_\mu +iA_\mu$ is a covariant
derivative.  They provide an interesting theoretical
laboratory. Indeed they present several features that hold in QCD:
Asymptotic freedom, gauge invariance, existence of a confining
potential between non gauge invariant states (that is eventually
screened by the dynamical constituents), and non-trivial topological
structure (instantons, $\theta$ vacua).  Moreover, unlike 
four-dimensional SU($N$)
gauge theories, a systematic $1/N$ expansion can be performed around
the large-$N$ saddle-point solution~\cite{DDL-79,Witten-79b,CR-92}.

Analogously to four-dimensional SU($N$) gauge theories, one may add a $\theta$
term to the Lagrangian, writing
\begin{equation}
{\cal L}_\theta  = {N\over 2g} \overline{D_\mu z}\, D_\mu z +
i \theta {1\over2\pi}\,\epsilon_{\mu\nu}\, \partial_\mu A_\nu,
\label{lagrangiancpntheta}
\end{equation}
where $q(x)={1\over2\pi}\,\epsilon_{\mu\nu}\, \partial_\mu A_\nu$ is
the the topological charge density.  Then one may study the $\theta$
dependence of the ground state and other observables.  In the
following we discuss this issue within the $1/N$ expansion, performed
keeping $g$ fixed.  Simple large-$N$ scaling arguments applied to the
Lagrangian (\ref{lagrangiancpn}) indicate that the relevant $\theta$
parameter in the large-$N$ limit should be $\bar{\theta}\equiv
\theta/N$.

Analogously to SU($N$) gauge theories, the ground state energy
$F(\theta)$ depends on $\theta$. One may define a scaling ground state
energy $f(\theta)$ and expand it around $\theta=0$,
\begin{equation}
f(\theta) \equiv M^{-2} [F(\theta)-F(0)] 
= {1\over 2} C \theta^2 \left( 1 + \sum_{n=1} b_{2n} \theta^{2n} \right) 
\label{fthetacpn} 
\end{equation}
where $F(\theta)$ is defined as in Eq.~(\ref{vftheta}), $M$ is the mass scale
at $\theta=0$ defined from the second moment of the two-point function of the
operator $P_{ij}(x) \equiv \bar{z}_i(x) z_j(x)$, $C$ is the scaling ratio
$\chi/M^2$ at $\theta=0$, where $\chi$ is the topological susceptibility. The
correlation function of the topological charge density, and in particular the
topological susceptibility, has been computed within the $1/N$ expansion
\cite{luescher-78,CR-91,v-99}.  We have
\begin{equation}
C = \chi/M^2 = {1\over 2\pi N} + O(1/N^2)
\end{equation} 
The coefficients $b_{2n}$ are obtained from appropriate $2n$-point
correlation functions of the topological charge density operators at
$\theta=0$.   The analysis of the $1/N$-expansion Feynman diagrams~\cite{CR-92} 
of the connected correlations necessary to compute $b_{2n}$
shows that they are suppressed in the
large-$N$ limit, as~\cite{DMPSV-06}
\begin{equation}
b_{2n} = O(1/N^{2n}).
\end{equation}
This implies that the ground-state energy can be rewritten as
\begin{eqnarray} 
&&f(\theta) = N \bar{f}(\bar{\theta}\equiv \theta/N), 
\label{fthetabarcpn}\\ 
&& \bar{f}(\bar{\theta}) = 
{1\over 2} \overline{C} \bar{\theta}^2 
( 1 + \sum_{n=1} \bar{b}_{2n} \bar{\theta}^{2n} ), 
\nonumber
\end{eqnarray}
where $\overline{C}\equiv N C$ and $\bar{b}_{2n}=N^{2n}b_{2n}$ are $O(N^0)$.
Note the analogy with the expected $\theta$ dependence
of the ground-state energy in SU($N$) gauge theories, cf.
Eq.~(\ref{fthetabar}).
Rather cumbersome calculations lead to the results~\cite{DMPSV-06}
$\bar{b}_2= - {27\over 5}$, and $\bar{b}_4= -{1830\over 7}$.

Within the $1/N$ expansion one may also study the dependence of the mass
$M$ on the parameter $\theta$. We write
\begin{equation}
M(\theta) = M\left( 1 + m_2 \theta^2 + ... \right)
\label{gmexcpn}
\end{equation} 
The analysis of its diagrams in the corresponding
$1/N$ expansion indicates that $m_2$ is suppressed as
\begin{equation}
m_2 = O(1/N^2)
\end{equation}
Once again, the relevant parameter is seen to be 
$\bar{\theta}\equiv \theta/N$\,.

\section{Critical slowing down of topological modes}
\label{slowdown}

Monte Carlo simulations of critical phenomena in statistical mechanics and of
quantum field theories, such as QCD, in the continuum limit are hampered by
the problem of critical slowing down (CSD) \cite{Sokal-92}.  The
autocorrelation time $\tau$, which is related to the number of iterations
needed to generate a new independent configuration, grows with increasing
length scale $\xi$.  In simulations of lattice QCD where the upgrading methods
are essentially local, it has been observed, see e.g. Refs.
\cite{ABDDV-96,DeFetal2,DPRV-02,DPV-02,LTW-04b} that the topological modes show
autocorrelation times that are typically much larger than those of other
observables not related to topology, such as Wilson loops and their
correlators.  Actually, the heating method~\cite{DV-92}, used to estimate
the topological susceptibility, essentially relies on this phenomenon.

Recent Monte Carlo simulations \cite{DPV-02,DPRV-02} of the
four-dimensional SU($N$) lattice gauge theories (for $N=3,4,6$) provided evidence of a
severe CSD for the topological modes, using a rather standard local
overrelaxed upgrading algorithm.  Indeed, the autocorrelation time $\tau_{\rm
  top}$ of the topological charge grows very rapidly with the length scale
$\xi\equiv \sigma^{-1/2}$, where $\sigma$ is the string tension, showing an
apparent exponential behavior $\tau_{\rm top}\sim \exp (c\xi)$ in the range of
values of $\xi$ where data are available.  Such a phenomenon worsens with
increasing $N$, indeed the constant $c$ appears to increase as $c\propto N$. 
Of course, this behaviour does not depend on the particular
estimator of the topological charge.  This peculiar effect has not been observed in
plaquette-plaquette or Polyakov line correlations, suggesting an approximate
decoupling between topological modes and nontopological ones, such as those
determining the confining properties.

These results suggest that the dynamics of the topological modes in Monte
Carlo simulations is rather different from that of quasi-Gaussian modes.  CSD
of quasi-Gaussian modes for traditional local algorithms, such as standard
Metropolis or heat bath, is related to an approximate random-walk spread of
information around the lattice.  Thus, the corresponding autocorrelation time
$\tau$ is expected to behave as $\tau\sim\xi^2$ (an independent configuration
is obtained when the information travels a distance of the order of the
correlation length $\xi$, and the information is transmitted from a given
site/link to the nearest neighbors).  This guess is correct for Gaussian (free
field) models; in general one expects that $\tau\sim \xi^z$ where $z$ is a
dynamical critical exponent, and $z\approx 2$ for quasi-Gaussian modes. On the
other hand, in the presence of relevant topological modes, he random-walk
picture may fail, and therefore we may have qualitatively different types of
CSD.  These modes may give rise to sizeable free-energy barriers separating
different regions of the configuration space.  The evolution in the
configuration space may then present a long-time relaxation due to transitions
between different topological charge sectors, and the corresponding
autocorrelation time should behave as $\tau_{\rm top}\sim \exp F_b$ where
$F_b$ is the typical free-energy barrier among different topological sectors.
However, this picture remains rather qualitative, because it does not tell us
how the typical free-energy barriers scale with the correlation length. For
example, we may still have a power-law behavior if $F_b \sim \ln \xi$, or an
exponential behavior if $F_b\sim \xi^\theta$.  It is worth mentioning that in
physical systems, such as random-field Ising systems \cite{Fisher-86} and
glass models \cite{Parisi-92}, the presence of significant free-energy
barriers in the configuration space causes a very slow dynamics, and an
effective separation of short-time relaxation within the free-energy basins
from long-time relaxation related to the transitions between basins.  In the
case of random-field Ising systems the free-energy barrier picture
supplemented with scaling arguments leads to the prediction that $\tau\sim
\exp (c \xi^\theta)$ where $\theta$ is a universal critical exponent
\cite{Fisher-86}.

The severe CSD experienced by the topological modes under local updating
algorithms should be a general feature of Monte Carlo simulations of lattice
models with nontrivial topological properties, since the mechanism behind this
phenomenon should be similar.  This has been also observed in two-dimensional
CP$^{N-1}$ models \cite{DMV-04,CRV-92}.  The numerical study of
Ref.~\cite{DMV-04} for various values of $N$ show that an exponential Ansatz,
i.e. $\tau_{\rm top}\sim \exp (c\xi^\theta)$ with $\theta\approx 1/2$,
and $c\propto N$,
provides a good effective description in the range of 
the correlation length $\xi$ where
data are available (however, the statistical analysis of the data
did not allow one to exclude an asymptotic
power-law behavior $\tau\sim \xi^z$
with $z\gtapprox N/2$ setting in at relatively large $\xi$).

The issue of CSD of topological modes is particularly important for lattice
QCD, because it may pose a serious limitation for numerical studies of
physical issues related to topological properties, such as the mass and the
matrix elements of the $\eta'$ meson, and in general the physics related to
the broken U(1)$_A$ symmetry. Indeed, it may substantially worsen the
cost estimates of the dynamical fermion simulations for lattice QCD, see,
e.g., Ref.~\cite{Jansen-03}.

Finally, we note that although the effects of the topological CSD have not been
directly observed in plaquette-plaquette or Polyakov line correlations, 
such a CSD
will eventually affect them.  The point is that the results of
Ref.~\cite{DMPSV-06}, summarized in Sec.~\ref{spectrum}, show that the
correlators of plaquette operators and topological charge do not vanish at
finite $N$, although they are quite small, and therefore there is not a
complete decoupling between topological and nontopological modes.  Therefore
the strong critical slowing down that is clearly observed in the topological
sector will eventually affect also the measurements of nontopological
quantities, such as those related to the string and glueball spectrum.

\end{document}